\documentclass{appolb}
\usepackage{graphicx} 

\usepackage[T1]{fontenc}
\usepackage[utf8]{inputenc}  

\usepackage{amsmath}
\usepackage{upgreek}  

\usepackage{cite}     
\usepackage{hyperref} 

\begin{document}
\title{Three-dimensional image reconstruction in J-PET using Filtered Back Projection method%
\thanks{Presented at the 2\textsuperscript{nd} Jagiellonian Symposium on Fundamental and Applied Subatomic Physics, Krak\'ow, Poland, June 4--9, 2017.}
}
\author{
R.~Y.~Shopa$^{a}$, K.~Klimaszewski$^{a}$, P.~Kowalski$^{a}$,  W.~Krzemie\'n$^{b}$, L.~Raczy\'nski$^{a}$, W.~Wi\'slicki$^{a}$, 
P.~Bia\l{}as$^c$, C.~Curceanu$^d$, E.~Czerwi\'nski$^c$, K.~Dulski$^c$, A.~Gajos$^c$, B.~G\l{}owacz$^c$, M.~Gorgol$^e$, B.~Hiesmayr$^f$, B.~Jasi\'nska$^e$, D.~Kisielewska-Kami\'nska$^c$, G.~Korcyl$^c$, T.~Kozik$^c$, N.~Krawczyk$^c$, E.~Kubicz$^c$, M.~Mohammed$^{c,g}$, M.~Pawlik-Nied\'zwiecka$^c$, S.~Nied\'zwiecki$^c$, M.~Pa\l{}ka$^c$, Z.~Rudy$^c$, N.~G.~Sharma$^c$, S.~Sharma$^c$, M.~Silarski$^c$, M.~Skurzok$^c$, A.~Wieczorek$^c$, B.~Zgardzi\'nska$^e$, M.~Zieli\'nski$^c$, P.~Moskal$^c$
\address{
$^{a}$ Department of Complex Systems, National Centre for Nuclear Research, 05-400 Otwock-\'Swierk, Poland \\
$^{b}$ High Energy Physics Division, National Centre for Nuclear Research, 05-400 Otwock-\'Swierk, Poland \\
$^{c}$ Faculty of Physics, Astronomy and Applied Computer Science, Jagiellonian University, 30-348 Cracow, Poland \\
$^{d}$ INFN, Laboratori Nazionali di Frascati, 00044 Frascati, Italy \\ 
$^{e}$ Institute of Physics, Maria Curie-Sk\l{}odowska University, 20-031 Lublin, Poland \\
$^{f}$ Faculty of Physics, University of Vienna, 1090 Vienna, Austria \\
$^{g}$ Department of Physics, College of Education for Pure Sciences, University of Mosul, Mosul, Iraq
}
}
\maketitle

\begin{abstract}
We present a method and preliminary results of the image reconstruction in the Jagiellonian PET tomograph. Using GATE (Geant4 Application for Tomographic Emission), interactions of the 511 keV photons with a cylindrical detector were generated. Pairs of such photons, flying back-to-back, originate from $e^+e^-$ annihilations inside a 1-mm spherical source. Spatial and temporal coordinates of hits were smeared using experimental resolutions of the detector. We incorporated the algorithm of the 3D Filtered Back Projection, implemented in the STIR and TomoPy software packages, which differ in approximation methods. Consistent results for the Point Spread Functions of $\sim5\div7$\,mm and $\sim9\div20$\,mm were obtained, using STIR, for transverse and longitudinal directions, respectively, with no time of flight information included.
\end{abstract}

\section{Introduction}

Recent studies of plastic scintillators \cite{Moskal2011, Moskal2014, Raczynski2014, Moskal2015, Raczynski2015, Kowalski2016, MoskalWieczorek2016, Raczynski2017, Smyrski2017} proved their usefulness in the Positron Emission Tomography (PET) for the detection of gamma-quanta that originate from the electron-positron $e^+e^-$ annihilation in matter. Such materials exhibit better timing properties than inorganic crystals, traditionally used in PET \cite{Conti2009, Humm2003, Karp2008, Townsend2004}, while low scintillation efficiency remains the main disadvantage. A novel Jagiellonian PET (J-PET) scanner, composed of plastic scintillator strips, utilizes Compton scattering for the detection of gamma-quanta \cite{MoskalWieczorek2016, Moskal2011, Moskal2014, Moskal2015}. This technology is expected to provide a superior figure of merit (see ref.~\cite{MoskalWieczorek2016}) for the whole body imaging, with good axial (longitudinal) and transaxial (transverse) resolution. Preliminary experiments, made for the 30-cm single scintillator strip, reveal temporal resolution of about 80\,ps (sigma of time of hit determination), corresponding to spatial resolution of $\sim$\,2.2\,cm (full width at half maximum -- FWHM) along axial direction \cite{Moskal2014}. Further tests are in progress now, for the already built full scale prototype of J-PET detector.

In this paper, we analyse the spatial resolution for the J-PET scanner, based on three-dimensional (3D) image reconstruction of a simulated \mbox{1-mm} spherical back-to-back gamma source, generated by the Geant4 Application for Tomographic Emission (GATE) simulation toolkit \cite{Gate2004, Gate2011}. Additional filtering and preprocessing were performed afterwards (described in ref.\,\cite{Kowalski2015,Kowalski2016}). Times and positions of interactions (hits) were smeared with Gaussian distributions, reflecting the experimental resolutions of the J-PET scanner.

The parameters of the simulations are suitable for the estimation of characteristics, defined by the National Electrical Manufacturers Association (NEMA) \cite{NEMA}. The norm NEMA-NU-2 requires 3D Filtered Back Projection (FBP) algorithm to be used for the image reconstruction. We therefore incorporated this method only, foreseeing the estimation of NEMA characteristics for J-PET. As no software for 3D FBP that correctly reflects J-PET geometry has been developed yet, we adapted methods from STIR \cite{STIR} and TomoPy \cite{TomoPy} packages. Both frameworks, however, impose constraints on the algorithm, which may impact the estimation of spatial resolution.

\section{Simulation setup}

\begin{figure}[htb]
\centerline{%
\includegraphics[width=9.8cm]{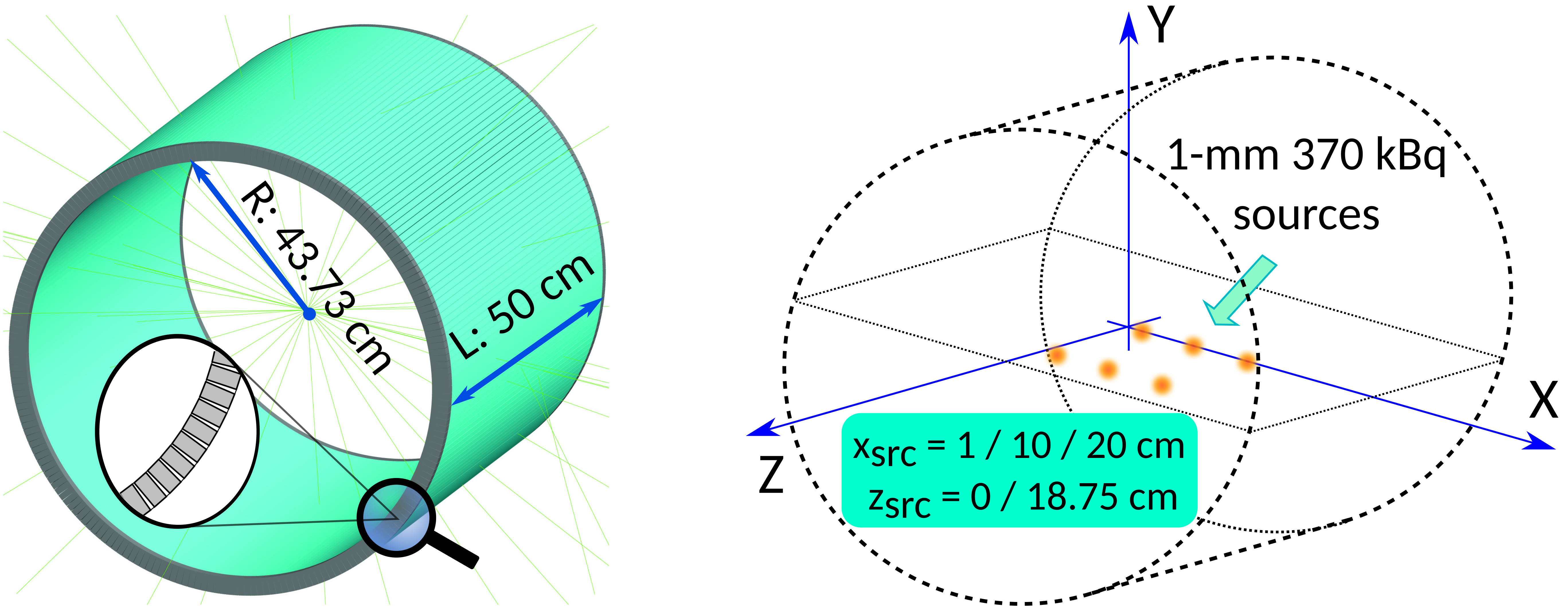}}
\caption{Left: simulated geometry of J-PET scanner with zoom depicting the composition of strips. Right: geometrical locations of the source, one per simulation.}
\label{Fig:JPETgeometry}
\end{figure}

Simulations in GATE framework \cite{Gate2004, Gate2011} were performed at the \'Swierk Computing Centre, National Centre for Nuclear Research. The J-PET scanner was defined for an ideal geometry (Fig.\,\ref{Fig:JPETgeometry}, left) with radius $R=437.3$\,mm and length $L=500$\,mm, comprising single layer of tightly composed plastic EJ-230 scintillator strips with rectangular cross-section and dimensions 7\,mm\,$\times$\,19\,mm\,$\times$\,500\,mm, 384 strips in total \cite{Kowalski2015, Kowalski2016}. 

A point-like source was defined as a simulated phantom. Its shape, size (a sphere 1\,mm in diameter) and activity (370\,kBq or 10\,$\upmu$Ci) were selected according to the NEMA requirements for the estimation of spatial resolution \cite{NEMA}. NEMA also defines locations of the source inside the scanner, specifically for detectors (like J-PET) that operate with large axial field of view (AFOV). Thus, we set 6 different, defined by the norm, locations $(x_{\text{src}}, y_{\text{src}}, z_{\text{src}})$ for the simulation (see Fig.\,\ref{Fig:JPETgeometry}, right). Along the axial coordinate $z$, the source was placed at the centre of the AFOV ($z_{\text{src}}=0$\,mm), or at the distance of three-eights of the AFOV length from the centre of the scanner ($z_{\text{src}}=187.5$\,mm). For the transverse direction, $x$ coordinate was defined at $x_{\text{src}}=10$\,mm, 100\,mm and 200\,mm with $y$ always at $y_{\text{src}}=0$\,mm.

Only subsample of simulated events (pairs of hits) was selected for the reconstruction, fulfilling selection criteria required for true coincidences, which include scattering filters and energy thresholds applied (see detailed description in refs.\,\cite{Kowalski2015, Kowalski2016}). At least 150\,000 events were used for the reconstruction, for each source position. The data include coordinates and times of hits inside the strips for each photon pair, i.e. $(x_1, y_1, z_1, t_1, x_2,y_2,z_2,t_2)$.

The tests, described in this article, encompasses three cases (for the smearing applied): assuming the spatial resolution in axial direction to be $\sigma_z=2$\,mm, 5\,mm or 10\,mm, whilst for the time of hit being fixed at $\sigma_t = 80$\,ps (corresponding to $\sigma_z=10$\,mm \cite{MoskalWieczorek2016}). The latter was chosen intentionally, to test the worst possible case of time of flight (TOF) taken into account.

Thus 3D sinogram could be composed, collecting lines of response (LOR) with TOF information optionally used for each event \cite{PETBasicScience}.

\section{Reconstruction procedure}

The reconstruction of the 3D image of the source, selected for the simulations, is essential for the estimation of spatial resolution of PET detectors, which is characterized by the so-called Point Spread Function (PSF). It is defined as the width of the reconstructed profile of a point source, measured similarly to FWHM along three principal axes \cite{PETBasicScience}.

\subsection{FBP algorithm} \label{FBP}
For a general 2D case, FBP algorithm, based on the inverse Radon transformation \cite{Helgason1984}, can be defined as mapping filtered sinogram $p^F(s,\phi)$ by an operator $X^*$, which returns an image $f(x,y)$:

\begin{equation}
f(x,y)=(X^*p^F)(x,y)= {\int_{0}^{\pi} d\phi p^F(s=x\cos\phi+y\sin\phi,\phi)}, \label{eq:FBP2D}
\end{equation}

\noindent where $p^F(s,\phi)$ is obtained by applying an apodized ramp filter $h(s)$ on the initial sinogram $p(s,\phi)$ \cite{PETBasicScience}:

\begin{equation}
p^F(s,\phi)= {\int_{-R_F}^{R_F} ds'p(s',\phi)h(s-s')}. \label{eq:RampFilter}
\end{equation}

\noindent Here $R_F$ denotes the radius of the field-of-view (FOV). 

In reality, scanner geometry requires sinogram variables (displacement $s$ and angle $\phi$) to be mapped onto discrete pairs $(s_i,\phi_i)$. Furthermore, it transforms (\ref{eq:FBP2D}) into a sum, containing functions $p^F_{i}(s=x\cos\phi+y\sin\phi,\phi_i)$. Here, $p^F_{i}(s,\phi_i)$ can be derived for any arbitrary Cartesian pair $(x,y)$ that defines unmapped $s$, using linear interpolation between $p^F(s_k,\phi_i)$ and $p^F(s_{k+1},\phi_i)$, calculated for two "known" neighbours $s_k$ and $s_{k+1}$ $(s_k<s<s_{k+1})$ \cite{PETBasicScience}. Practical implementations of FBP occasionally incorporate such interpolation for the increase of the resolution of the reconstructed image $f(x,y)$.

Equations for 3D FBP could be derived by turning values into vectors in (\ref{eq:FBP2D})-(\ref{eq:RampFilter}) and simplifying the formulas, eventually using 4-dimensional sinogram functions $p(s,\phi,\zeta,\theta)$, where $\zeta$ and $\theta$ denote axial coordinate and tilt angle (obliqueness) of LOR plane, respectively \cite{PETBasicScience}.

\subsection{Software frameworks}
For the reconstruction of the point source using 3D FBP, required by \cite{NEMA}, STIR \cite{STIR} and TomoPy \cite{TomoPy} software packages were chosen.

\begin{figure}
\centerline{%
\includegraphics[width=10.5cm]{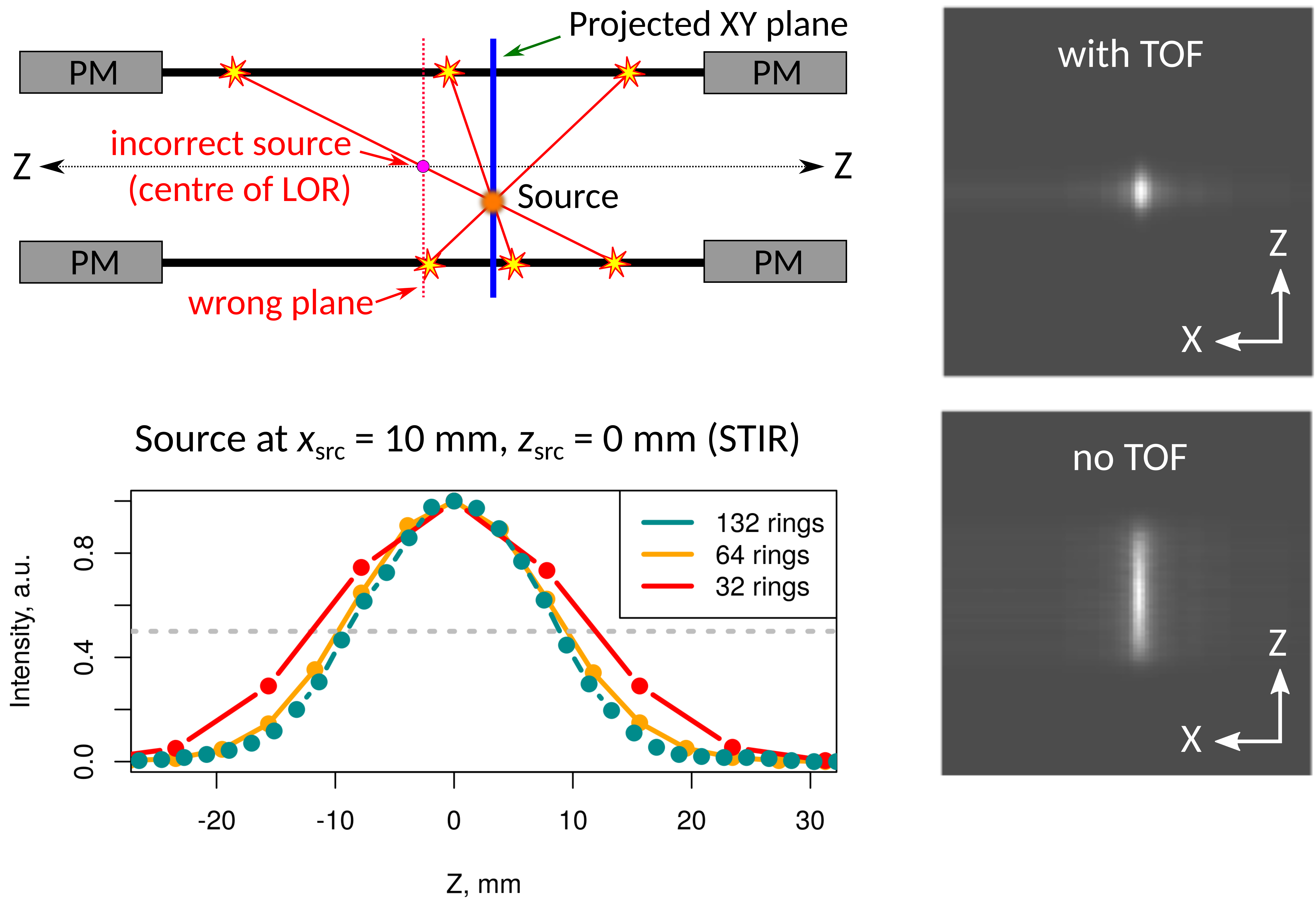}}
\caption{Left top: LOR projection scheme. Transverse planes with and without TOF corrections are denoted by solid blue and dashed red lines, respectively. Right: reconstructed image in TomoPy (using 2D FBP ASTRA algorithm) for both cases with $Z$ axis aligned vertically. Left bottom: axial intensity profile of the reconstructed image for various strip discretisation by rings.}
\label{Fig:TOFTomoPy}
\end{figure}

Since TomoPy does not provide full 3D version of the FBP algorithm, additional data transformation must be applied. All LORs are projected onto $XY$-planes, orthogonal to $Z$ axis (as in Fig.\,\ref{Fig:TOFTomoPy}, left top), which would also reduce the size of 3D sinogram. Reconstructed image is a stack of $XY$-slices, each obtained using 2D FBP ASTRA algorithm \cite{PeltASTRA}. The number of such slices is not limited, but the simplified procedure would constrain the estimation of the longitudinal spatial resolution $\text{PSF}_z$. However, it could evidently be improved by adjusting axial coordinate of projected planes to $e^+e^-$ annihilation point along LOR, estimated from times of hits $t_1$ and $t_2$, which reflect TOF data (see Fig.\,\ref{Fig:TOFTomoPy}). It is important to note that such "TOF correction" does not change FBP algorithm, with no additional kernel used, unlike known TOF reconstruction methods (see, for example, \cite{AnnTOF}).

STIR accepts few input formats with Interfile type of data storage \cite{CradduckInterfile} used for reconstruction. Because of that, for conversion of list mode output from GATE to Interfile we employed a dedicated package, developed for SAFIR project \cite{SAFIR2017, SAFIRgithub}. Unlike TomoPy, STIR conserves the LOR geometry in sinogram completely, including its obliqueness towards $Z$ axis. Moreover, for the compensation of partial truncation of the image near the edges of AFOV, the Kinahan and Rogers method \cite{Kinahan1990} for 3D FBP is implemented.

STIR and SAFIR define cylindrical scanner as a stack of axially aligned narrow rings, made of smaller size detectors. Therefore, $z$-component of each hit position is mapped to discrete values and eventually stored as a 1-byte index, which sets the limit of maximum number of rings to 256 \cite{SAFIRgithub}. This might distort the image during reconstruction, deteriorating the $\text{PSF}_z$ estimation for J-PET. We observed little change, though, in intensity profile along $Z$ direction, for the number of rings above 64 (Fig.\,\ref{Fig:TOFTomoPy}, left bottom).

TomoPy determines transverse resolution of the reconstructed image (number of pixels) from FOV discretisation, i.e. from the number of J-PET strips in our case. For better analysis, we additionally incorporated PET package \cite{PETPackage}, developed for 2D image reconstruction in R environment \cite{Rcran}. Unlike TomoPy, both STIR and PET package have an option to increase image resolution for $X$ and $Y$ axes, using interpolation technique, mentioned in sect.\,\ref{FBP}, which allows more accurate PSF estimation.

\section{Results and discussion}

The spatial resolution $\text{PSF}_{x,y,z}$ of J-PET was determined by FWHM estimation in transverse ($X,Y$) and longitudinal ($Z$) directions from the reconstructed images of a point source. Two cuts, made along maximum intensity of $Z$- and $Y$-axis ($XY$ and $XZ$ planes, respectively), were created for TomoPy and STIR (only $XY$ for R PET package). Pixel size of each cut determines the error of $\text{PSF}_{x,y,z}$. For $X$ and $Y$ directions, these errors are the same with maximum value of $\Delta x_{\text{TomoPy}}=3.6$\,mm. Applying interpolation, we diminished transaxial pixel size to $\Delta x_{\text{STIR}}=1.8$\,mm (by setting the zoom parameter in STIR to 2) and $\Delta x_{\text{R PET}}<0.9$\,mm. 

\begin{figure}
\centerline{%
\includegraphics[width=12.5cm]{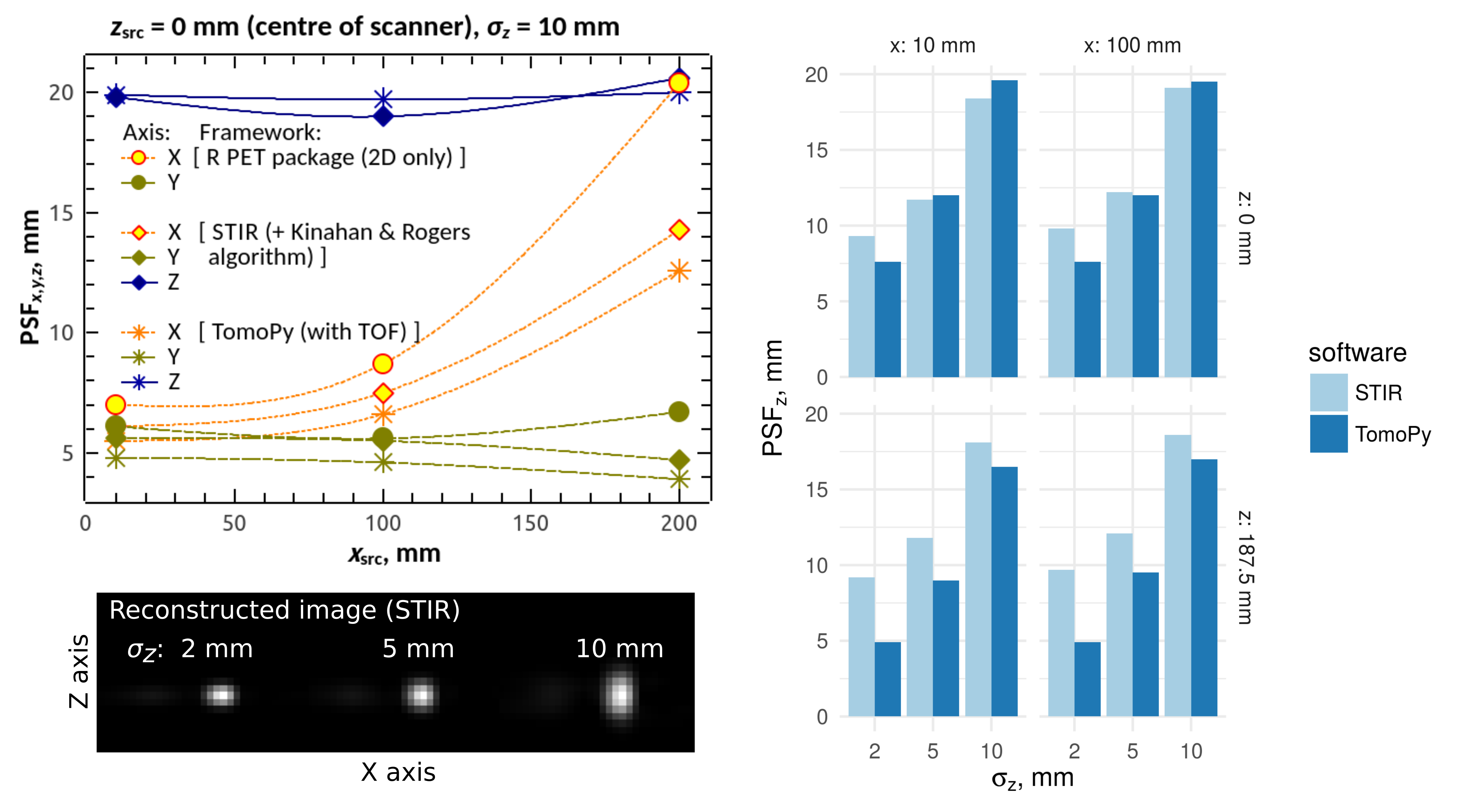}}
\caption{Left top: preliminary results for $\text{PSF}_{x,y,z}(x_{\text{src}})$ dependencies, estimated in different software frameworks, for $z_{\text{src}}=0$\,mm and $\sigma_z = 10$\,mm. Left bottom: $XZ$-projections of reconstructed images in STIR, for different smearing parameter $\sigma_z$. Right: preliminary estimations of longitudinal $\text{PSF}_z$ for various positions of the source and smearing parameters $\sigma_z$ (notation "src" omitted for $x$ and $z$).}
\label{Fig:BigResults}
\end{figure}

The longitudinal errors for TomoPy (200\,cuts along $Z$) and STIR (64 rings) were $\Delta z_\text{TomoPy}=2.5$\,mm and $\Delta z_\text{STIR}=3.9$\,mm, respectively. Such parameters are sufficient for the PSF estimation for different $\sigma_z$ (see left bottom of Fig.\,\ref{Fig:BigResults}). Resulting voxel size ($\Delta x\times\Delta y\times\Delta z$) for STIR and TomoPy was $1.8$\,mm\,$\times1.8$\,mm\,$\times3.9$\,mm and $3.6$\,mm\,$\times3.6$\,mm\,$\times2.5$\,mm, respectively.

For the transverse direction, the values of $\text{PSF}_{x,y}$ that correspond to various frameworks, tend to differ significantly if the source is distant from the centre of the scanner (Fig.\,\ref{Fig:BigResults}, left top). In comparison, $\text{PSF}_{x,y}$ are inside the span of $\Delta x$ for the position, closest to the centre ($x_{\text{src}}=10$\,mm, $y_{\text{src}}=0$\,mm, $z_{\text{src}}=0$\,mm), resulting in effective spatial resolution of $\sim 5\div7$\,mm. We did not observe any influence of $\sigma_z$, reflecting the axial smearing of hit position, on the transverse resolution: in all cases $\text{PSF}_{x,y}$ were increasing similarly if moving phantom far from the centre.

Estimated longitudinal resolution $\text{PSF}_z$ differs for various software used. TOF correction, implemented in TomoPy, reduces $\text{PSF}_z$ systematically for the case $z_{\text{src}}=187.5$\,mm, compared to $z_{\text{src}}=0$\,mm (darker bars in Fig.\,\ref{Fig:BigResults}, right). The explanation of this is the activity of the source and irradiation time: if the source is closer to the edge, much smaller obliqueness is allowed for LORs (see Fig.\,\ref{Fig:TOFTomoPy}), so that both photons would interact with the scanner. Therefore, longer exposure is required for $z_{\text{src}}=187.5$\,mm to achieve $\sim$150\,000 events that match selection criteria. Together with TOF correction it would definitely improve resolution. Conversely, Kinahan and Rogers algorithm, implemented in STIR, artificially adds oblique LORs by expanding scanner along $Z$, which eventually diminishes the role of the exposure \cite{Kinahan1990}.

To sum up, implementation of TOF technique proved to be promising for J-PET scanners. Minimal PSF, estimated in TomoPy for the smallest smearing $\sigma_z=2$\,mm is below 5\,mm in both longitudinal and transverse directions, which is comparable to modern, commercially available TOF-PET/CT systems \cite{EJNMMI2016, Slomka2016}. However, these results are yet to be confirmed by more accurate approach, since full 3D FBP algorithm has not been employed. For the correct estimation, all NEMA requirements are to be fulfilled. 

\begin{table}[htb]
\centering
\caption{Estimated (preliminary) PSF values for 3D FBP reconstruction of \mbox{1-mm} point source in STIR framework, located at $(x_\text{src}, y_\text{src}, z_\text{src})$ in J-PET scanner. Transaxial position is defined by $x_\text{src}$ ($y_\text{src}$ is equal to zero), "centre" and "edge" denote axial coordinate ($z_{\text{src}}=0$\,mm and $z_{\text{src}}=187.5$\,mm, respectively). Voxel size $\Delta x\times\Delta y\times\Delta z=1.8$\,mm\,$\times1.8$\,mm\,$\times3.9$\,mm.} 
\vspace{0.2cm}
\begin{tabular}{cccccccc}
\hline
\hline \\[-1.9ex]
&&\multicolumn{6}{c}{PSF,\,mm} \\
\multicolumn{2}{r}{Along axis:} & \multicolumn{2}{c}{$X$} & \multicolumn{2}{c}{$Y$} & \multicolumn{2}{c}{$Z$} \\
\hline
$\sigma_z$,\,mm & $x_{\text{src}}$,\,mm & centre & edge & centre & edge & centre & edge \\
\hline \\[-1.5ex]

& 10 & 5.8 & 5.8 & 5.8 & 5.4 & 9.3 & 9.2 \\
2.0 & 100 & 7.7 & 7.6 & 5.4 & 5.4 & 9.8 & 9.7 \\
\vspace{0.1cm}
& 200 & 13.8 & 13.8 & 4.5 & 4.5 & 9.7 & 9.7 \\

& 10 & 5.8 & 5.8 & 5.8 & 5.4 & 11.7 & 11.8 \\
5.0 & 100 & 7.1 & 7.1 & 5.4 & 5.4 & 12.2 & 12.1 \\
\vspace{0.1cm}
& 200 & 13.4 & 13.4 & 4.5 & 4.5 & 13.6 & 12.6 \\

& 10 & 5.8 & 5.8 & 5.8 & 5.4 & 18.4 & 18.1 \\
10.0 & 100 & 7.6 & 7.6 & 5.4 & 5.4 & 19.6 & 16.5 \\
\vspace{0.1cm}
& 200 & 13.4 & 13.8 & 4.5 & 4.5 & 19.4 & 20.4 \\

\hline  
\end{tabular}
\label{Tab:STIRFWHM2}
\end{table}

On the contrary, STIR 3D FBP algorithms return consistent and reproducible results. Hence, preliminary PSF values, obtained in this framework, should be considered reliable, with projected longitudinal resolution for \mbox{J-PET} scanner of $\sim9\div10$\,mm and $\sim18\div19$\,mm for the smallest and the largest $\sigma_z$, respectively, with no TOF correction applied. Summarized results are listed in Table\,\ref{Tab:STIRFWHM2}. As one can see, there is little influence of axial position of the source on the longitudinal resolution, as well as axial smearing on the transverse resolution. This might be considered as a benefit of STIR application in J-PET for 3D image reconstruction.

\section*{Acknowledgements}

We acknowledge STIR/SAFIR technical support of PhD Jannis Fischer (Department of Physics, Institute for Particle Physics, ETH Z\"urich, Switzerland) and the support by the National Science Centre through the grant No.\,2016/21/B/ST2/01222.


\end{document}